\documentclass[a4paper]{article}

\usepackage{INTERSPEECH2019}

\usepackage{etoolbox}
\usepackage[dvipsnames]{xcolor}

\usepackage{url} 

\newcommand{\accessedDate}{May~2020}
\newcommand{\urlfootnote}[1]{\footnote{\url{#1} Accessed \accessedDate}}

\title{An Open source Implementation of ITU-T Recommendation P.808 with Validation}
\name{Babak Naderi$^1$, Ross Cutler$^2$}
\address{
  $^1$Quality and Usability Lab, Technische Universit\"at Berlin\\
  $^2$Microsoft Corp.}
\email{babak.naderi@tu-berlin.de, ross.cutler@microsoft.com}

\begin{document}

\maketitle

\begin{abstract}
The ITU-T Recommendation P.808 provides a crowdsourcing approach for conducting a subjective assessment of speech quality using the Absolute Category Rating (ACR) method. 
We provide an open-source implementation of the ITU-T Rec. P.808 that runs on the Amazon Mechanical Turk platform. 
We extended our implementation to include Degradation Category Ratings (DCR) and Comparison Category Ratings (CCR) test methods. We also significantly speed up the test process by integrating the participant qualification step into the main rating task compared to a two-stage qualification and rating solution. We provide program scripts for creating and executing the subjective test, and data cleansing and analyzing the answers to avoid operational errors.
To validate the implementation, we compare the Mean Opinion Scores (MOS) collected through our implementation with MOS values from a standard laboratory experiment conducted based on the ITU-T Rec. P.800. 
We also evaluate the reproducibility of the result of the subjective speech quality assessment through crowdsourcing using our implementation.
Finally, we quantify the impact of parts of the system designed to improve the reliability: environmental tests, gold and trapping questions, rating patterns, and a headset usage test.
\end{abstract}

\section{Introduction}

Assessment of the quality of the transmitted speech via a telecommunication system, the so-called Quality of Experience (QoE) \cite{moller2014quality}, has been the subject of research since the early days of telephony. Due to the subjective nature of the QoE, the speech quality is commonly assessed by test participants in either Listening-opinion or Conversation-opinion tests. During the subjective test, participants either listen to a short speech file (listening-opinion) or hold a conversation using the system under the test (conversation-opinion) and then give their opinion about the perceived quality on one or more rating scales. These tests are typically carried out by following the ITU-T Recommendation P.800 \cite {ITU-P800} in a laboratory environment to control the influence of all external factors on the participants' judgments.
Although controlled laboratory setup increases the reliability of measurement, it lacks realism as the listening system, and the test environment does not reflect the typical usage situations. Objective measures of speech quality such as PESQ, SRMR, and P.563 have been shown to have a low correlation to listening opinion, and are not reliable replacements to listening opinion tests even though they are widely used by researchers due to their convenience \cite{avila2019}. 

Meanwhile, micro-task crowdsourcing offers a fast, low-cost, and scalable approach to collect a subjective assessment from a geographically distributed pool of demographically diverse participants using a diverse set of listening devices.
In the crowdsourcing subjective test, the assessment task should be provided over the Internet to the crowdworkers who carry out the tasks using their own devices and in their working environment. They are typically remunerated for their work. 
The ITU-T Rec. P.808 \cite{ITU-P808} is a recently published standard that describes how to conduct a subjective evaluation test of speech quality using the crowdsourcing approach.

The recommendation details the crowdsourcing listening-opinion test by specifying the experiment design, test procedure, and data analysis for the Absolute Category Rating (ACR) method. It contains descriptions for online test methods that evaluate participant's eligibility (i.e. hearing test), environment and listening system suitability, and also quality control mechanisms. A previous study showed that the ITU-T Rec. P.808 provides a valid and reliable approach for speech quality assessment in crowdsourcing \cite{naderi2020impact}.
We provide an open-source implementation\urlfootnote{https://github.com/microsoft/P.808} of the ITU-T Rec. P.808 to avoid misinterpretations, operational errors and ultimately to make subjective speech quality assessment accessible for the entire research and industry community without a huge cost of building a dedicated test laboratory. The implementation includes usage documentation, program scripts for creating and executing the crowdsourcing subjective test, data cleansing and analyzing the answers in compliance with ITU-T Rec. P.808. 

The paper is organized as follows: In Section 2, we discuss related work. In Section 3, we describe our implementation and the extensions we provided. The results of validation tests are reported in Section 4. There we focus on the validity and reproducibility of assessment provided by our implementation. A discussion and proposals for future work conclude the paper in Section 5.
\section{Related work}
There are several open source crowdsourcing systems for estimating subjective speech quality, which are summarized in Table~\ref{table:1}. 
Some solutions require separate servers to use, and some like ours can be run directly on crowdsourcing platforms like Amazon Mechanical Turk (AMT) which makes the solution more convenient to use.  
Hearing and environmental tests help make subjective tests more accurate, and are part of lab-based studies.
Validation means the system has been tested and compared against a lab-based study and shown to be accurate. 
Reproducibility means that system has been tested and shown to give consistent results at different times with different raters.
Our solution is the first that meets all of these requirements.
\begin{table}[h!]
\begin{center}
\scalebox{.65}{
\begin{tabular}{ c c c c c c c  } 
\toprule
 \textbf{Tool} & \textbf{Measures} & \textbf{Server} & \textbf{Hearing} & \textbf{Envir.} & \textbf{Validation} & \textbf{Repro.} \\
 & & & \textbf{test} & \textbf{test} \\
  \midrule
 CrowdMOS \cite{ribeiro2011crowdmos} & ACR, MUSHRA & Y & N & N & Y & Y (N=2) \\
 QualityCrowd2 \cite{Keimel2012}  & ACR & Y & N & N & N & N \\
 BeaqleJS \cite{beaqlejs} & ABX, MUSHRA & Y & N & N & N & N \\
 WESP+ \cite{Rainer2013} & ACR, DCR, PC, & Y & N & N & N & N \\
  &  DSIS, DSCQS & & & \\
 webMUSHRA \cite{schoeffler_webmushra_2018} & AB, ABX, & Y & N & N & N & N \\
  & MUSHRA, ACR &  &  &  \\
 CAQE \cite{Cartwright2018} & MUSHRA, PC & N & Y & N & Y & N\\
 P.808 (ours) & ACR, DCR, CCR & N & Y & Y & Y & Y \\
\bottomrule
\end{tabular}
}
\end{center}
\caption{Comparison of open source audio subjective quality systems}
\label{table:1}
\end{table}

\section{P.808 implementation}
Our implementation uses the internal structure of the Amazon Mechanical Turk\urlfootnote{https://mturk.com}. Therefore there is no need for a dedicated webserver for conducting the test. However, the experimenter needs to provide URLs of the speech clips. In the following tools provided by P.808 implementation and the extensions we did are described.


\subsection{Tools}
To avoid operational errors and ease the interaction with the system, we provide program scripts for creating the subjective test (e.g., create customized trapping questions dataset), post-processing the answers (i.e., data cleansing, reconstructing and aggregating the ratings) and for interacting with AMT (e.g., sending bonuses to workers or notifying them).

\begin{figure}[htbp]
\centerline{\includegraphics[width=\columnwidth]{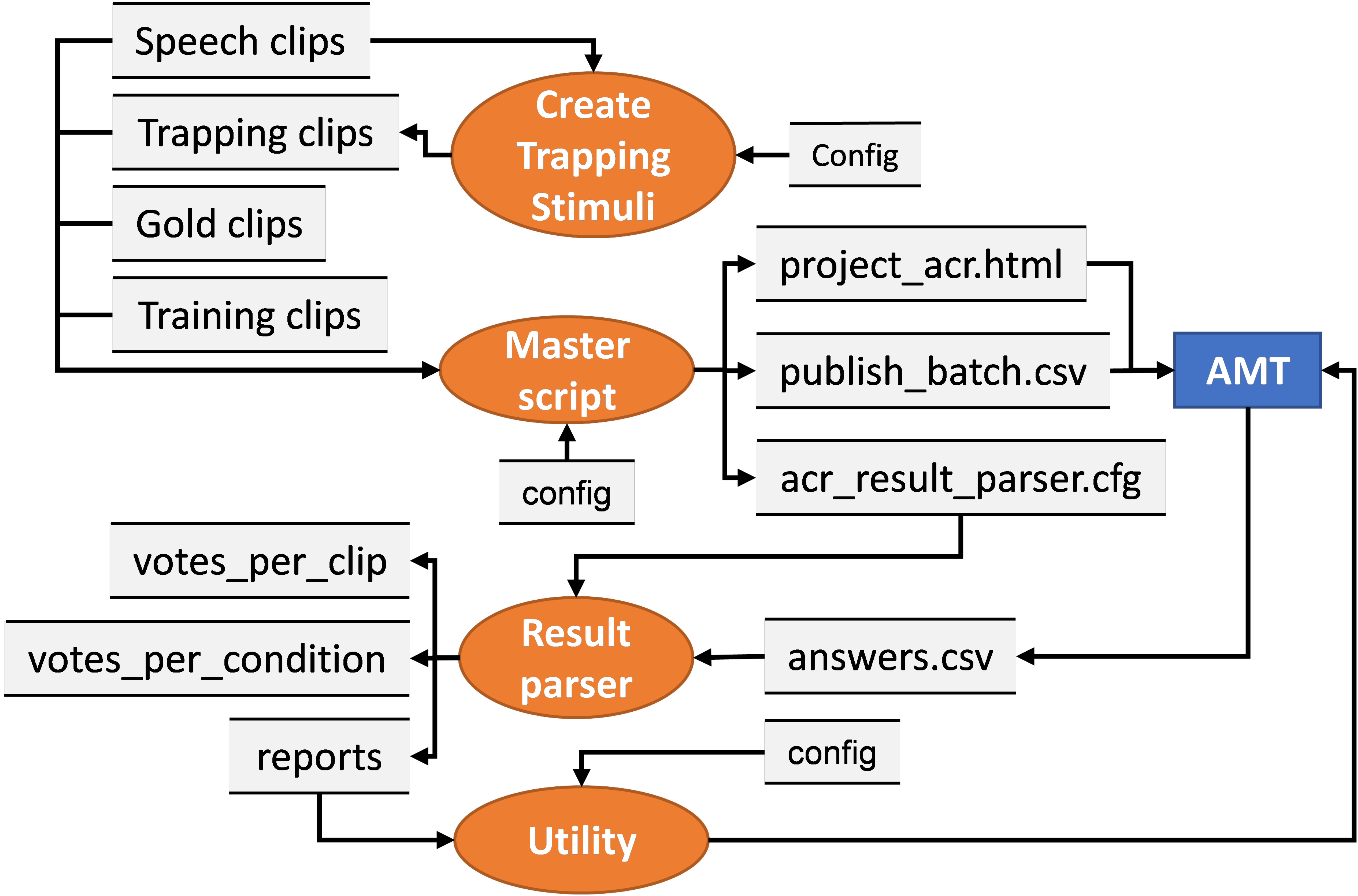}}
\caption{Data Flow Diagram. }
\label{fig_dfd}
\end{figure}

\begin{figure}[htbp]
\centerline{\includegraphics[width=1\columnwidth]{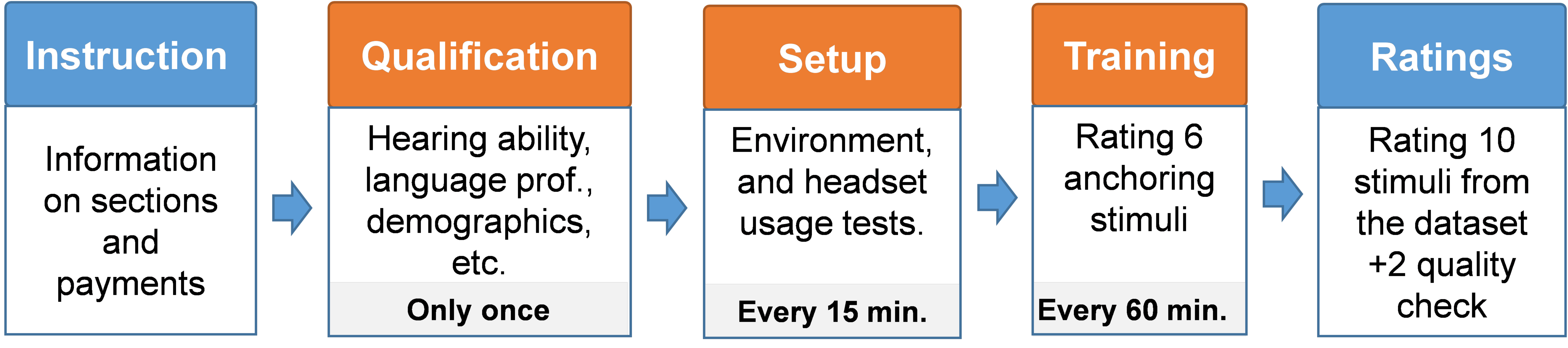}}
\caption{Sections of the P.808 assignment. Intervals and number of stimuli can be changed in the configuration file.}
\label{fig_ssections}
\end{figure}

The master script (c.f. Figure\ref{fig_dfd}) generates the HTML file (also known as a HIT App), input URLs, and the configuration file for the result parser (i.e. post-processing script), given the URLs of all clips in the dataset, training stimuli, trapping stimuli, and gold-standard stimuli as inputs.
The generated HTML file contains the hearing test, environment suitability test \cite{naderi2020env}, usage of two-eared headphones check \cite{polzehl2015robustness}, and trapping and gold standard questions \cite{naderi2015effect} as specified in the ITU-T Rec. P.808. The stages of the crowdsourcing task from the worker's perspective are illustrated in Figure~\ref{fig_ssections}.

The post-processing script marks a submitted answer accepted when all audio clips in the session are fully played, the worker successfully passed the test for correct usage of both earpods, and the trapping question is correctly answered. However, it only uses the ratings provided in the accepted answer package in further steps when the environmental suitability test is satisfied, the gold standard question is correctly answered, and there is enough variance in the ratings.
Beside data-cleansing report, the post-processing script generates utility reports (i.e., accepted, rejected submission, and assigned bonuses), reconstructs and aggregate the reliable ratings per stimuli and per test condition (if it exists) and provides typical statistics, i.e., Mean Opinion Score (MOS), standard deviations, number of valid votes, and 95\%~ Confidence Interval

\subsection{Extensions}
The open source implementation provides an exact implementation of the ITU-T Rec. P.808 and also extends it in different aspects. The general extensions speed up the test and increase its validity. We also implemented the Degradation Category Ratings (DCR) and Comparison Category Ratings (CCR) test procedures which we suggest to be used for extending the ITU-T Rec. P.808.

\subsubsection{Integrated rater qualification}
The recommendation introduces a multi-step job design in which workers first perform a qualification task; based on their response a randomly selected group of workers, who are eligible to participate in the test, will be invited to the training or directly to the rating job.

The qualification job should examine crowdworker's hearing ability, language proficiency, and ask the type of listening device they have, their demographics, and if they have been directly involved in work connected to speech quality. 
Although the multi-step design guarantees that only workers who satisfied the prerequisites get access to the rating job, it strongly slows down the rating-collection process as it depends on the availability of a limited group of workers. In addition, in platforms like AMT with many job offers, the rating job might get lost between all other available tasks. Consequently, there is also no guaranty that workers who finally provide the ratings represent the target demography.

We integrated the worker qualification as a section into the rating task. 
The qualification section will be shown only once to the worker as a part of the first rating task. Their response will directly be evaluated, and the result will be saved in their web-browser's local storage. Despite their performance, the worker can submit their answer and will be compensated for their time. In case they successfully passed the qualification, this section will be invisible in the next assignment. Otherwise, the entire task will be disabled, and the worker is informed that there is no more assignment that matches their profile.
The post-processing script also checks for eligibility of worker's qualifications. It also rejects fraud responses i.e., using other browsers or manipulating the values stored in the web browser's local storage.
The integration of rater qualification into the rating job made the entire process scalable and reduced the execution time by 4-5X in several test runs.

\subsubsection{Headset detection}
The ITU-T Rec. P.808 urges participants to use a two-eared headphone/headset. Previous works showed that listening through loudspeakers leads to smaller discrimination capacity than headphones \cite{ribeiro2011crowdmos}. It is because their judgment is more influenced by their surrounding acoustic scene.
However, no online screening method is proposed rather than asking participants in the qualification job and only inviting those who self-reported having a headset.
We use the API of the WebRTC standard to get a list of all known connected audio devices to the user's device. Parsing the device names may reveal if there is a headset connected is used. 
Currently, on average, we could detect a headset device in 40.1\% of sessions.
The results of this method should be interpreted with caution, as not detecting a headset does not imply that the user does not have a connected headphone. The WebRTC API does not list a wide variety of devices without microphones.

\subsubsection{Periodic environment test}
Within the environment test, participants should listen to 4 pairs of stimuli and select which one has better quality. It reveals whether the working environment is suitable enough for participating in the test at that specific point of time \cite{naderi2020env}. As the rating section typically contains 12 stimuli only, listening to 8 stimuli for environment tests every time adds considerable overhead. 
We introduce a temporal environment suitability certificate. When the certificate does not exist, the environment-test section will be injected into the rating assignment. Participant's answers will be evaluated simultaneously. In case they successfully passed the test, a temporal certificate will be stored in the local storage of the web browser. A temporary certificate that expires in 30 minutes has reduced the overall working time of participants by $40\%$.

\subsubsection{Extensions in test procedures}
We provide the implementation of the DCR and CCR test procedures for crowdsourcing by following their description for laboratory experiments according to the ITU-T Rec. P.800 and the structure of the ACR test in crowdsourcing. Their main differences compared to the ACR procedure are that participants should listen to two stimuli (one is the reference stimulus, and the other is the processed stimulus), and provide their opinion on different scales. In the DCR procedure, participants first listen to the reference stimulus and then the processed one. They rate their degree of annoyance on the five-point degradation category scale (from \textit{1. Degradation is very annoying} to \textit{5. Degradation is inaudible}). In the CCR procedure, the reference stimulus is not revealed to the participant, and the order of presentations is randomized. The participant rates the quality of the second stimulus compared to the quality of the first stimulus on a 7-point scale (from \textit{-3: Much Worse} to \textit{3.Much Better}).
In each rating session, we inject a trapping question in which both stimuli are the reference one.
The post-processing script considers the presentation order and calculates the DMOS (Degradation Mean Opinion Score) and CMOS (Comparison Mean Opinion Score) for DCR and CCR procedures, respectively.

\section{Validation}
\subsection{Validity}
\begin{figure}[htbp]
\centerline{\includegraphics[width=0.7\columnwidth]{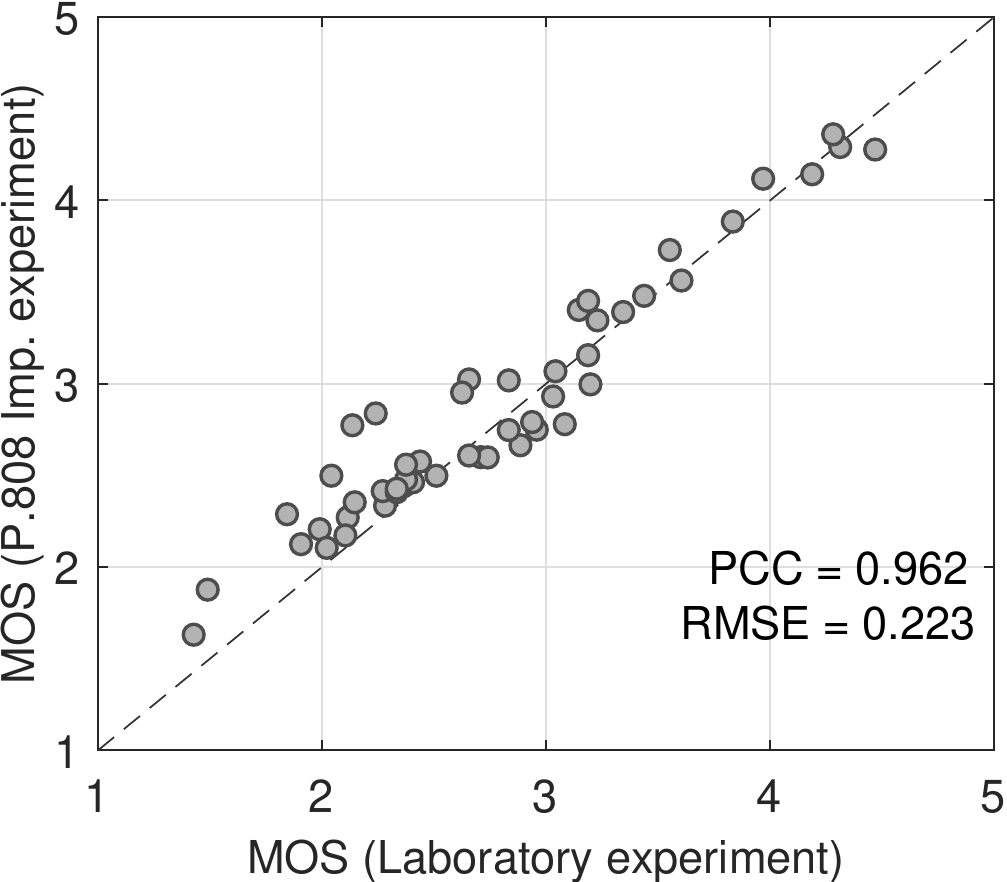}}
\caption{Comparison between MOS from our P.808 implementation and MOS from P.800 Laboratory study (dataset: ITU-T Sup.23).}
\label{fig_scatter}
\end{figure}

To estimate the validity of our P.808 system we compare P.808 results with a rated dataset that has been evaluated in a lab environment using the ITU-T Rec. P.800, namely ITU Supplement 23 \cite{ITU-Psup23}. This supplement is an evaluation of the narrowband G.729 codec. We used an English dataset from the ``Experiment 3 Effect of Channel Degradations'' in the Supplement 23 study, which includes 200 files in 50 conditions of channel degradation. In the lab study, 96 votes per condition were collected. 
We conducted a study using our P.808 implementation on the same dataset. We have collected on average 260 valid votes per condition. We randomly selected 96 votes per condition and calculate the MOS values. The Pearson Correlation Coefficient (PCC), and Spearman's Rank Correlation Coefficient (SRCC), are computed using the MOS results for each condition number with respect to the lab-based P.800 MOS results reported in ITU Supplement 23. We repeated the sampling ten times and on average we achieved PCC = 0.954, SRCC = 0.923, and RMSE = 0.237 which reduced to 0.214 after first-order mapping. Figure \ref{fig_scatter} illustrates the MOS values of each condition number in the lab and our p.808 study (using all votes) in a scatter plot.

\subsection{Reproducibility}
\begin{figure}[htbp]
\centerline{\includegraphics[width=.9\columnwidth]{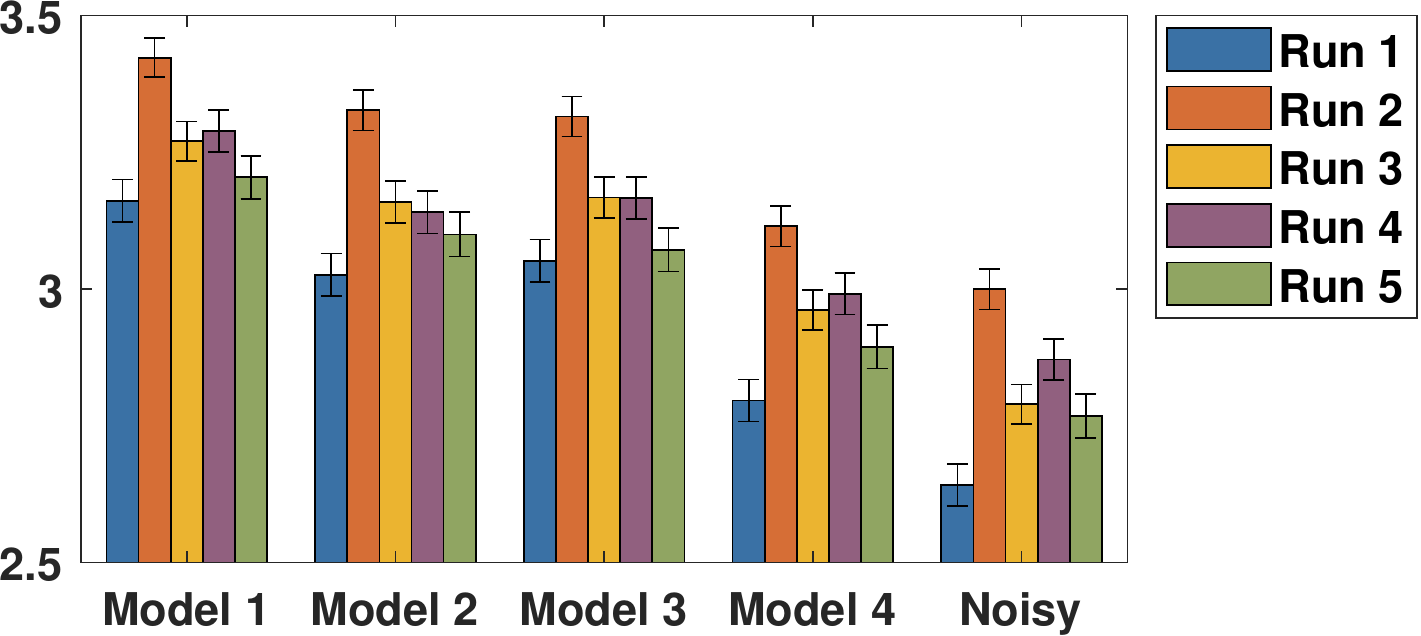}}
\caption{MOS per test conditions in the reproducibility study.}
\label{fig_rep1}
\end{figure}

To determine how reproducible our P.808 system is we conducted a study on the wideband dataset used in the INTERSPEECH Deep Noise Suppression Challenge \cite{reddy2020interspeech}. We used the 700 clip test set and ran P.808 five times on five separate days, each day with unique raters. On average 89 workers participated in each run. For this study we followed the ACR with the Hidden Reference test method, i.e., we had five test conditions, four deep noise suppression models, and the noisy speech which is the hidden reference. We targeted to collect 5 votes per stimulus. The data-screening process of the P.808 implementation on average approved 88.71\% of submission using its default filtering criteria. The calculated Differential Mean Opinion Score (i.e., $DMOS=MOS_{M}-MOS_{noisy}$ where $M$ is a test condition) values are reported in Table~\ref{tab:reproducibility}. On average we observed $PCC=0.994$ and  $SRCC=0.94$ between MOS values of models in the five different runs. Figure~\ref{fig_rep1} shows the MOS values for five models\footnote{Model 5 is the test condition with reference signals (noisy).} and error bars represent the 95\% CI. The bias observed between MOS values in different runs is a well-known and common behavior in the subjective tests \cite{ITU-P1401}. Using the DMOS successfully removed that offset. In case more conditions were under tests, a typical solution is to use anchor test conditions in the test and apply first- or third-order mapping for removing the bias \cite{ITU-P1401}.

Applying the transformation proposed by Naderi and Möller \cite{naderi2020transformation} leads to $SRCC=1.00$. We also calculate the Intraclass Correlation Coefficient (ICC) to examine the reliability of the P.808 implementation based on single measures, absolute agreement, and two-way random model. It shows how strongly each run of the P.808 implementation resembles each other.
Results show a good reliability when considering MOS values ($ICC = 0.719$) and an excellent reliability when using DMOS ($ICC=0.907$).

\begin{table}[t]
    \caption{DMOS values observed in the reproducibility study using ACR-HR approach.}
    \label{tab:reproducibility} 
    \begin{center}
        \begin{tabular}{ c c  c  c  c  c }
        \toprule
        \textbf{Name} &	\textbf{Run1}&	\textbf{Run2}&	\textbf{Run3}&	\textbf{Run4}&	\textbf{Run5} \\ 
        \midrule
        Model1&	0.52&	0.42&	0.47&	0.43&	0.43 \\ 
        Model2&	0.37&	0.32&	0.36&	0.28&	0.33 \\ 
        Model3&	0.40&	0.31&	0.36&	0.30&	0.31 \\ 
        Model4&	0.16&	0.11&	0.17&	0.13&	0.14 \\ 
        \bottomrule
        \end{tabular}
    \end{center}
\end{table}


\subsection{Accuracy improvements}
We used the above-mentioned dataset to examine how each integrated filter enhances the reproducibility and reliability of the results of the P.808 implementation.
We divided the submission into two groups, i.e., the submissions which pass a filter and the submissions which failed due to that filter. We calculated the MOS values for each test condition based on ratings in each group and examined for those two groups how consistent the resulting scores are in the five different runs.
We only considered the criteria that enough submissions were failed due to them being able to calculate meaningful MOS values based on the failed submissions, namely gold stimulus, environment test, headset test, and all filters together. 
We calculated ICC, average PCC, and average SRCC over five runs (Table~\ref{tab:performance}). 
We also evaluated the significance of the difference between correlation coefficients from the passed and the failed groups using Fisher-z transformation, as suggested in \cite{ITU-P1401}.

Results show that submissions that passed the integrated filters are more consistent in reproducing similar results. Significant improvements were observed when the gold stimulus, environment test (despite $ICC_{MOS}$), and all criteria together applied. None of the criteria leads to a significant difference in SRCC, and the headset filter did not lead to a significant improvement. As mentioned before, the headset detection test only was successful in 40\% of cases and therefore is not conclusive.

It should be noted that the number of submissions in the passed and failed group was not equal, and the ratio differed between filtering criteria. Therefore we cannot conclude the exact impact of each filter on the reliability of subjective measurements through crowdsourcing using this data. Instead, reported results should be considered as an example of such an effect on a typical use case. For instance, only eight submissions failed because of the wrong answer to the trapping questions in all five runs\footnote{We included a sample trapping question in the training section for the participants. Consequently, the number of submission failing because of a wrong answer to the trapping questions in rating procedure significantly reduced.}. It cannot be concluded that the trapping question component is not beneficial as previous studies showed that even its presence has a significant impact on participants' behavior and encouraging them to provide higher quality responses \cite{naderi2015effect, naderi2015effectobs}.
\begin{table}[t]
\caption{Effect of integrated filtering criteria on reliability of P.808 implementation.}
\label{tab:performance} 
\begin{center}
\resizebox{\columnwidth}{!}{%
    \begin{tabular}{lp{6mm}p{6mm}p{6mm}p{6mm}p{6mm}p{6mm}p{6mm}p{6mm}}
    \toprule
    \textbf{Criteria} & \multicolumn{2}{c}{\textbf{ICC\textsubscript{MOS}}} & \multicolumn{2}{c}{\textbf{ICC\textsubscript{DMOS}}}    & \multicolumn{2}{c}{$\overline{ \textbf{PCC}}$} & \multicolumn{2}{c}{$\overline{\textbf{SRCC}}$} \\
    
     & \textit{\footnotesize Passed} & \textit{\footnotesize Failed} & 
     \textit{\footnotesize Passed} & \textit{\footnotesize Failed}  & 
     \textit{\footnotesize Passed} & \textit{\footnotesize Failed}  &  
     \textit{\footnotesize Passed} & \textit{\footnotesize Failed}  \\
    \midrule

    Gold stimulus  & \textbf{.705}\textsuperscript{*} & \textbf{.063} & \textbf{.899}\textsuperscript{*} & \textbf{.281} &    \textbf{.995}\textsuperscript{*} & \textbf{.824} & .94 & .74\\
    Env. Test      & .685 & .451 & \textbf{.901}\textsuperscript{*} & \textbf{.455} &	\textbf{.994}\textsuperscript{*} & \textbf{.855}   & .94 & .71\\
    All criteria   & \textbf{.719}\textsuperscript{a} & \textbf{.286} & \textbf{.907}\textsuperscript{*} & \textbf{.556} &	\textbf{.994}\textsuperscript{*} & \textbf{.893} & .94 & .82\\
    \midrule
    Headset det.  & .588 & .605 & .825 & .818 &	.981 & .992  &	 .94 & .94\\
    
    \bottomrule
    \multicolumn{5}{l}{$^{*1}$ Significant at $\alpha=.05$}
    \end{tabular}%
}
\end{center}
\end{table}


\section{Conclusions}
We have provided an open source implementation of the ITU-T Rec. P.808 and validated that it correlates well with a lab study and the results are reproducible. We have enhanced the implementation to significantly reduce test execution time and added DCR and CCR test procedures as well. We have used the P.808 implementation on dozens of subjective tests internally and shown it to be an essential tool for our audio processing development. It has also been used in the INTERSPEECH Deep Noise Suppression challenge \cite{reddy2020interspeech}, which made the challenge possible to do with similar quality as a lab study but with much lower costs and effort to administrate. The cost is low enough that it can even be part of a daily machine-learning development cycle. 

There are several extensions to the tool for future work. For speech enhancement work, we'd like to include a template for ITU-T Rec. P.835 \cite{ITU-P835}, which would provide separate MOS values for the speech signal, background noise, and overall quality. This would provide more detailed subjective listing metrics, and ultimately lead to further improvement of the speech enhancement machine learning models. Also, instead of periodic assessment of the test environment, we would like to have continuous monitoring of the environment (e.g., estimating the SPL) and triggering a new environment test when the acoustic scene has significantly degraded. Finally, we would like to make the ratings on specific runs be extensible with other runs, using anchoring conditions, a more strict environment test, and first- or third-order mapping in the post-processing step.


\bibliographystyle{IEEEtran}

\bibliography{mybib}

\end{document}